\documentclass[12pt,preprint]{aastex}

\newcommand{\eq}{\begin{equation}}
\newcommand{\ee}{\end{equation}}

\def\t0{\theta_{\circ}}

\def\be{\begin{equation}}
\def\en{\end{equation}}

\def\gapp{\ \lower 3pt\hbox{${\buildrel > \over \sim}$}\ }
\def\lapp{\ \lower 3pt\hbox{${\buildrel < \over \sim}$}\ }

\shorttitle{Optical Afterglow Observations of GRB 040924}
\shortauthors{Huang et al.}

\begin{document}

\title{Optical Afterglow Observations of the Unusual
Short-Duration Gamma-Ray Burst 040924}

\author{
K. Y. \textsc{Huang}\altaffilmark{1},
Y. \textsc{Urata}\altaffilmark{2,3},
A. V. \textsc{Filippenko}\altaffilmark{4},
J. H. \textsc{Hu}\altaffilmark{1},
W. H. \textsc{Ip}\altaffilmark{1},
P. H. \textsc{Kuo}\altaffilmark{1},
W. \textsc{Li}\altaffilmark{4},
H. C. \textsc{Lin}\altaffilmark{1},
Z. Y. \textsc{Lin}\altaffilmark{1},
K. \textsc{Makishima}\altaffilmark{2,5},
K. \textsc{Onda}\altaffilmark{2,6},
Y. \textsc{Qiu}\altaffilmark{7},
and T. \textsc{Tamagawa}\altaffilmark{2}
}

\altaffiltext{1}{Institute of Astronomy, National Central University,
Chung-Li 32054, Taiwan, Republic of China (Huang email: d919003@astro.ncu.edu.tw).}  
\altaffiltext{2}{RIKEN (Institute of
Physical and Chemical Research), 2-1 Hirosawa, Wako, Saitama 351-0198,
Japan.}  
\altaffiltext{3}{Department of Physics, Tokyo Institute of
Technology, 2-12-1 Ookayama, Meguro-ku, Tokyo 152-8551, Japan.}
\altaffiltext{4}{Department of Astronomy, University of California,
Berkeley, CA 94720-3411.}
\altaffiltext{5}{Department of Physics,
University of Tokyo, 7-3-1 Hongo, Bunkyo-ku, Tokyo 113-0033, Japan.}
\altaffiltext{6}{Department of Physics, Tokyo University of Science,
1-3 Kagurazaka, Shinjyuku-ku, Tokyo, Japan.}
\altaffiltext{7}{National Astronomical Observatories, Chinese Academy
of Sciences, Beijing 100012, China, PR.}  


\begin{abstract}

The 1-m telescope at Lulin Observatory and the 0.76-m Katzman
Automatic Imaging Telescope at Lick Observatory were used to observe
the optical afterglow of the short-duration (1.2--1.5~s) gamma-ray
burst (GRB) 040924. This object has a soft high-energy spectrum, thus
making it an exceptional case, perhaps actually belonging to the
short-duration tail of the long-duration GRBs. Our data, combined with
other reported measurements, show that the early $R$-band light curve
can be described by two power laws with index $\alpha = -0.7$ (at $t =
16$--50 min) and $\alpha = -1.06$ (at later times). The rather small
difference in the spectral indices can be more easily explained by an
afterglow model invoking a cooling break rather than a jet break.

\end{abstract}


\section{Introduction}

Gamma-ray bursts (GRBs) are among the most powerful explosions in
the Universe. It is generally believed that the impulsively injected
fireball results from core collapse in a massive star (Woosley 1993;
MacFadyen \& Woosley 1999), or from the merging of either two neutron
stars or a neutron star and a black hole (e.g., Ruffert et al. 1997;
Popham et al. 1999; Narayan et al. 1992, 2001; Rosswog \& Davies
2002; Lee \& Ramirez-Ruiz 2002). After the explosion, the relativistic
ejecta collide with the ambient interstellar medium causing
X-ray, optical, and radio emission. These so-called ``afterglows'' thus
carry important information on the injection mechanism, the
configuration of the (possibly collimated) fireball, and the
surrounding environment (e.g., M\'esz\'aros 2002).

Two kinds of GRBs have been defined according to whether their
gamma-ray emission has duration longer or shorter than 2~s. Although
their frequency distributions overlap, that of the short-duration GRBs
peaks at 0.3~s, while that of the long-duration GRBs peaks at 30--40~s
(Kouveliotou et al.  1993). In addition, the duration is weakly
correlated with the spectral hardness ratio at high energies: short
GRBs tend to be harder and long GRBs tend to be softer (Kouveliotou et
al.  1993).

The optical afterglows of short GRBs were elusive until the
detection (Fenimore et al. 2004) of GRB 040924 by the High Energy
Transient Explorer 2 ({\it HETE--2}) on 2004 Sep. 24, at 11:52:11 (UT
dates are used throughout this paper). This event lasted about 1.2~s
and was X-ray rich according to the {\it HETE--2} flux in the
7--30~keV and 30--400~keV bands. The Konus-Wind satellite also
detected this event with 1.5~s duration in the 20--300~keV band
(Golenetskii et al. 2004). Since the high-energy spectrum of GRB
040924 is soft (Fenimore et al. 2004), the object might actually be
near the short-duration end of the long GRBs. A detailed study of the
associated optical afterglow could provide further information on
whether this is indeed the case, thus probing the nature of GRBs in
the boundary region.

About $t = 16$~min after the burst, Fox (2004) detected the
corresponding optical afterglow with an $R$-band magnitude of
$\sim$18. This was shortly followed by Li et al. (2004), who reported
$R \approx 18.3$ mag at 26~min after the burst. From then on, a number
of observatories joined in the follow-up observations (Fynbo et
al. 2004; Hu et al. 2004; Khamitov et al. 2004a; Terada, Akiyama, \&
Kawai 2004). Radio observations failed to detect the afterglow at $t =
12.54$~hr and $t = 5.79$~d (Frail \& Soderberg 2004; van de Horst,
Rol, \& Wijers 2004a,b). Spectral measurements by the Very Large
Telescope (VLT) of a galaxy located at the position of the optical
afterglow indicated a redshift $z = 0.859$ for this event (Wiersema et
al. 2004).

\section{Observations and Data Analysis}

Upon receiving the burst alert message from {\it HETE--2} and the
optical position reported by Fox \& Moon (2004), multi-band (Johnson
$B$ and $V$; Bessell $R$ and $I$) follow-up observations of GRB 040924
with the Lulin One-meter Telescope (LOT, in Taiwan) were initiated according
to the previously approved Target-of-Opportunity procedure. Photometric
images were obtained with the PI1300B CCD camera ($1300 \times 1340$
pixels, $\sim 11' \times 11'$ field of view; Kinoshita et al. 2005)
during the interval 14.31--20.89 on Sep. 24 (i.e., 2.4--9.0~hr after
the burst). Unfortunately, the earliest observations ($t < 3.1$ hr)
were not successful because of poor weather and short exposure
times. These problems also affected all of the $B$ and $I$ data, and
many of the $V$ and $R$ images as well. Nevertheless, our observations
reveal unusual early-time evolution of the afterglow brightness, as
discussed below.

A standard routine including bias subtraction and flat-fielding
corrections with appropriate calibration data was employed to process
the data using IRAF.\footnote[8]{IRAF is distributed by the National
Optical Astronomy Observatory, which is operated by AURA, Inc., under
cooperative agreement with the National Science Foundation.}  The
afterglow was clearly seen in the $V$-band and $R$-band images (Figure
1). The position of the afterglow is $\alpha$(J2000) $= 02^{\rm h}
06^{\rm m} 22^{\rm s}.52$, $\delta$(J2000) $= +16^{\circ}06'48''.82$
($\pm 0''.23$ in each coordinate).  Next, the DAOPHOT package (Stetson
1989) was used to perform aperture photometry of the GRB field by
choosing ten field stars for differential photometry. The LOT data
were combined with median filtering to improve the signal-to-noise
ratio. For the photometry, the aperture diameter was set to 4 times
the FWHM of the objects. The magnitude error was estimated as
$\sigma_{\rm e}^{2}= \sigma_{\rm ph}^{2} + \sigma_{\rm sys}^{2}$,
where $\sigma_{\rm ph}$ is the photometric error of the afterglow
estimated from the DAOPHOT output, and $\sigma_{\rm sys}$ is the
systematic calibration error estimated by comparing the instrumental
magnitudes of the ten field stars. Besides the calibration data
obtained by the USNOFS 1.0-m telescope (Henden 2004), we used the
measurements of four Landolt (1992) standard-star fields (SA92,
PG2331+055, SA95, and PG2317+046) taken by LOT on a photometric
night. The difference between the two flux calibrations is within 0.04
mag. The magnitudes derived for the $R$ and $V$ observations are
summarized in Table 1.

In addition to the LOT data, we have also included two early
measurements from the 0.76-m Katzman Automatic Imaging Telescope
(KAIT; Li et al. 2003b) at Lick Observatory at $t = 0.43$ and
1.06~hr. The KAIT data were taken without filters, but the
transformation of the unfiltered magnitude to $R$ can be determined
from the $V-R$ colors of the GRB field stars and of the optical
afterglow (Li et al. 2003a,b). The calibration of the GRB 040924 field
is adopted from Henden (2004), and the value $V-R = 0.57$ mag of the
afterglow is taken from LOT data at 0.292~d after the burst. KAIT
observed the GRB at low airmass (1.26--1.4), and the local standard
stars have $V-R$ colors (0.39--0.85 mag) similar to that of the
GRB. Moreover, from three photometric nights we found that the
coefficient for the second-order extinction is only 0.04; thus, the
errors caused by second-order extinction of GRB 040924 are small, and
are included in the overall uncertainties of the KAIT data.

\section{Results}

The light curve of GRB 040924 in Figure 2 is a combination of the
early observations reported by Fox (2004) at 0.012~d and 0.033~d after
the burst, the work reported here, and the measurements by Khamitov et
al. (2004b,c) at $t = 0.37$~d, 0.62~d, and 1.56~d, Fynbo et al. (2004) at
$t = 0.73$~d, and Silvey et al. (2004) around $t = 0.9$~d. To put all of
the data onto a consistent magnitude scale, we recalibrated the
above-mentioned published data by using the Henden (2004) standard
stars for the GRB 040924 field. The data of Fox (2004) were calibrated
by GSC~2.2 stars\footnote[9]{The GSC~2.2 is a magnitude-selected
subset of GSC~II, an all-sky catalog based on $1''$ resolution scans
of the photographic Sky Survey plates, at two epochs and three
bandpasses, from the Palomar and UK Schmidt telescopes
(http://www-gsss.stsci.edu/gsc/gsc2/GSC2home.htm).}  with F-emulsion
magnitude which corresponds closely to the $R$-band magnitude; the GSC
stars are 0.11 mag brighter than the Henden standard stars in the
average of our images. Since two reference stars are provided by
Khamitov et al. (2004b,c) and Fynbo et al. (2004), we measured these
stars from LOT $R$-band combined images and obtained the average magnitudes
and root-mean-square errors; the results were then used to recalibrate
the reported afterglow magnitudes.

The time evolution of the light curve can be expressed in terms of a
power law with $F(t) \propto t^{\alpha}$, where $t$ is the time after
the burst and $\alpha$ is the index. We find $\alpha = -0.87 \pm 0.02$
($\chi^{2}/\nu$= 0.06 for $\nu=2$) for the very sparse $V$-band data (only
three closely spaced LOT observations and one later observation from Silvey
et al. 2004). Similarly, we derive $\alpha = -0.99 \pm 0.02$ ($\chi^{2}/\nu$
= 2.08 for $\nu=12$) from all 14 available $R$-band observations.
These two values of $\alpha$ fall within the range of long GRBs
($\alpha = -0.62$ to $-2.3$), so the afterglow of GRB 040924 is
consistent with the standard model of cosmic-ray electrons accelerated
by the internal and external shocks of the expanding fireball
\citep{meszaros}, as in the case of typical long-duration GRBs.

Upon closer scrutiny, the first three $R$-band observations (at $t = 16$--50
min) indicate $\alpha = -0.7$, consistent with the conclusion of Fox (2004),
while the subsequent data (starting from the third $R$-band observation)
give a somewhat steeper value of $\alpha = -1.06 \pm 0.03$ (with $\chi^{2}/\nu = 1.09$ for
$\nu = 10$). [Essentially the same late-time result, $\alpha = -1.06 \pm 0.02$, is found
when we use only our own LOT and KAIT data, together with the Sibley et al.
(2004) observation at $t = 0.91$~d.] The data thus suggest
the presence of a mild break, the significance of which is discussed
below.

Finally, our LOT observations of GRB 040924 at $t = 7.10$~hr indicate
a color index of $V-R = 0.57 \pm 0.18$ mag, corrected for foreground
reddening of $E(B-V) = 0.058$ mag (Schlegel, Finkbeiner, \& Davis
1998).  We have also calculated the color of observations by Silvey et
al. (2004) at $t = 0.91$~d (22.1~hr) to be $V-R = 0.35 \pm 0.10$ mag,
corrected for foreground reddening. While these two values are consistent
with the color of typical long GRBs ($V-R = 0.40 \pm 0.13$ mag; Simon et
al. 2001), they also may suggest the interesting possibility of a color
change during the time evolution of this afterglow. However, the color
change is only marginally significant, given the uncertainties. Future GRB
afterglow measurements should shed new light on this tantalizing behavior.

\section{Discussion}

The brightness variations of the optical afterglows of GRBs potentially
yield important information on the expansion of the ejecta. For example,
breaks in the light-curve power laws at several hours to several days after
the bursts have been observed in a number of GRBs. This effect is generally
believed to be associated with the evolution as a collimated jet (Rhoads 1999).
In the case of GRB 040924, because of the small variation from $\alpha =
-0.7$ to $\alpha = -1.06$ around $t = 50$~min, the break is not well
constrained. On the other hand, this small break could be indicative of some
interesting physical process. Note that the amplitude of the break
($\Delta\alpha = \alpha_2 - \alpha_1$; here $\alpha_1$ and $\alpha_2$ are
the power-law indices before and after the break, respectively) is independent
of extinction under the assumption of no color change. In the case of GRB
040924, from $\alpha_1 = -0.7$ and $\alpha_2 = -1.06$ we find $\Delta\alpha
\approx -0.36$. According to theoretical work (Rhoads 1999), $\Delta\alpha
= -3/4$ for a collimated jet with a fixed angle, and $\Delta\alpha = \alpha_1
/3 -1 \approx -1.23$ for a sideways-expanding jet in the framework of a constant
ambient density model. It is clear that the amplitude of the break in GRB
040924 is much smaller than values expected of jet expansion with power-law
indices much steeper after the break. The interpretation of a jet break for
GRB 040924 is thus uncertain. Next we will explore an alternative possible
explanation.

Panaitescu \& Kumar (2000) pointed out that a light-curve break could
also be caused by the spectral cooling frequency moving through the
optical band. This property might be used as a diagnostic tool to
differentiate among different possible scenarios of GRB afterglow
formation. In the standard GRB afterglow model, it is usually assumed
that the synchrotron emission observed in optical bands originates
from the expansion of a blast wave of constant energy into an
interstellar medium (ISM) of constant density. However, there is also
increasing evidence that some GRBs have massive-star progenitors.
Consequently, the corresponding relativistic blast waves should actually
be expanding into the stellar wind of the progenitor stars with a density
variation of $\rho \propto r^{-2}$ (Dai \& Lu 1998; M\'esz\'aros et
al. 1998; Panaitescu et al. 1998).  Zhang \& M\'esz\'aros (2004)
listed the broad-band optical spectra of the synchrotron radiation from
a power-law distribution of energetic electrons with a spectral index
($p$) accelerated by the blast wave; accordingly, we can obtain the
values of $p$ before and after the cooling break.

As shown in Table 2, the ISM model provides the only possible fit (for
$p > 2$) to the GRB 040924 observations which satisfies the
requirement that $p$ should remain nearly the same ($p \approx 1.93$
to 2.08) as the spectrum evolves from $\nu_{\rm opt} < \nu_{\rm c}$ to
$\nu_{\rm opt} > \nu_{\rm c}$. Note that within the framework of the
ISM model, $p = 2.33$ for $\nu_{\rm opt} < \nu_{\rm c}$ and $p = 2.00$
for $\nu_{\rm opt} > \nu_{\rm c}$ if the corresponding light curve can
be characterized by a single power-law index ($\alpha = -0.99 \pm
0.02$).

Another interesting estimate can be made concerning the relation
between the cooling-break frequency $\nu_{\rm c}$, the break time
$t_{\rm day}$ (in units of days), the redshift $z$, the magnetic
energy $\varepsilon_{\rm B}$, the kinetic energy $E_{52}$ (in units of
$10^{52}$ erg), and the density $n_0$ of the ISM. According to Granot
\& Sari (2004),
\begin{equation}
\nu_{\rm c} =
6.37(p-0.46)10^{13}e^{-1.16p}(1+z)^{-1/2}\varepsilon_{\rm B}^{-3/2}n_0^{-1}E_{52}^{-1/2}t_{\rm day}^{-1/2} .
\end{equation}

\noindent
Now, with $t_{\rm day} = 0.035$, $\nu_{\rm c} = 4.7 \times 10^{14}$~Hz
in the $R$ band, $z = 0.859$, $p \approx 2.08$, and the assumptions
that $E_{52} = 1.48$ and $n_0 = 1$ cm$^{-3}$, we find
$\varepsilon_{\rm B} \approx 0.16$. This value is consistent with the
normal assumption for the magnetic-energy fraction of
$\varepsilon_{\rm B} \approx 0.1$, though slightly larger. For the
case of a single power-law index ($\alpha = -0.99 \pm 0.02$),
$\varepsilon_{\rm B} < 0.01$, which is much smaller than the normal
value. Our analysis thus suggests that the observed light curve of GRB
040924 could be the result of the spectral cooling frequency moving
through the optical band. In other words, the apparent break in GRB
040924 might not be a jet break but rather a cooling break.

\section{Conclusion}

The 1.2--1.5~s duration of GRB 040924, though formally in the domain
of short GRBs, overlaps the short end of long-duration GRBs. Moreover,
it has a soft high-energy spectrum, characteristic of long GRBs. Our
optical afterglow observations show that the temporal evolution,
power-law index, and $V-R$ color of GRB 040924 are also consistent
with those of well-observed long GRBs. The signature of a
low-amplitude break in the light curve, as suggested by our present
data, can be explained by the afterglow model invoking a cooling break
at early times. However, note that the jet break usually
occurs 1--2~d after the burst, and there are few observations of
GRB 040924 at $t > 1$~d. Thus, we cannot exclude the possibility
that the true jet break occurred outside the range of our
observations.

Due to the general lack of information on the optical afterglows of
short GRBs, we cannot compare our observations of GRB 040924 to this
class of objects. The $\it{Swift}$ satellite, with higher gamma-ray
sensitivity and more accurate localization than previous missions,
will provide more opportunities to understand the properties of
typical short GRBs and of GRBs near the boundary between short and
long GRBs.

\bigskip

We thank the staff and observers at the Lulin telescope for various
arrangements that made possible the observations reported herein. This
work is supported by grants NSC 93-2752-M-008-001-PAE and NSC 93-2112-M-008-006.
Y.U. acknowledges support from the Japan Society for the Promotion of
Science (JSPS) through a JSPS Research Fellowship for Young Scientists.
A.V.F. is grateful for NSF grant AST-0307894, and for a Miller Research
Professorship at UC Berkeley during which part of this work was completed.
KAIT was made possible by generous donations from Sun Microsystems, Inc.,
the Hewlett-Packard Company, AutoScope Corporation, Lick Observatory, the
NSF, the University of California, and the Sylvia and Jim Katzman Foundation.

\clearpage

\begin{figure}
\plotone{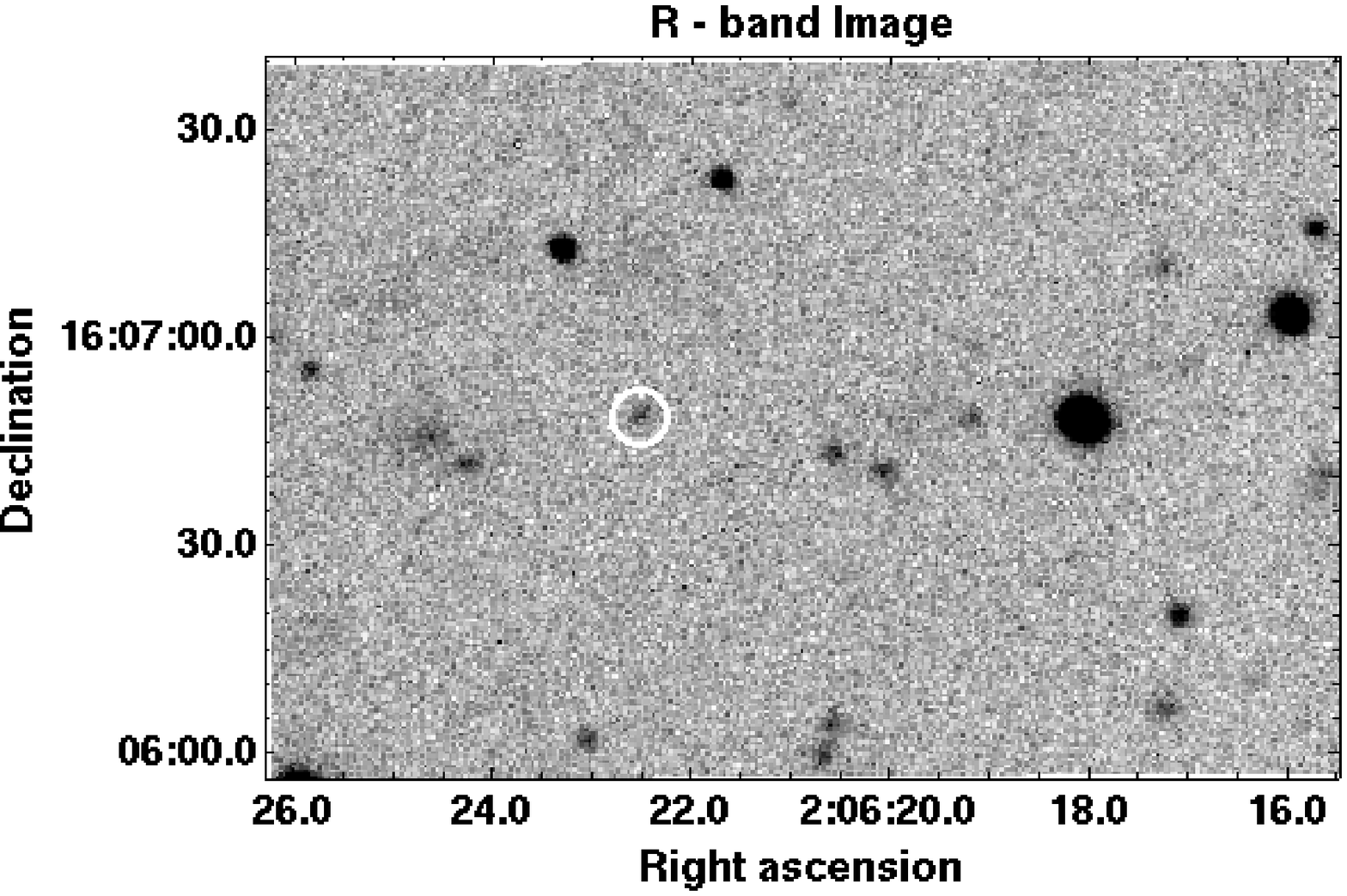}
\plotone{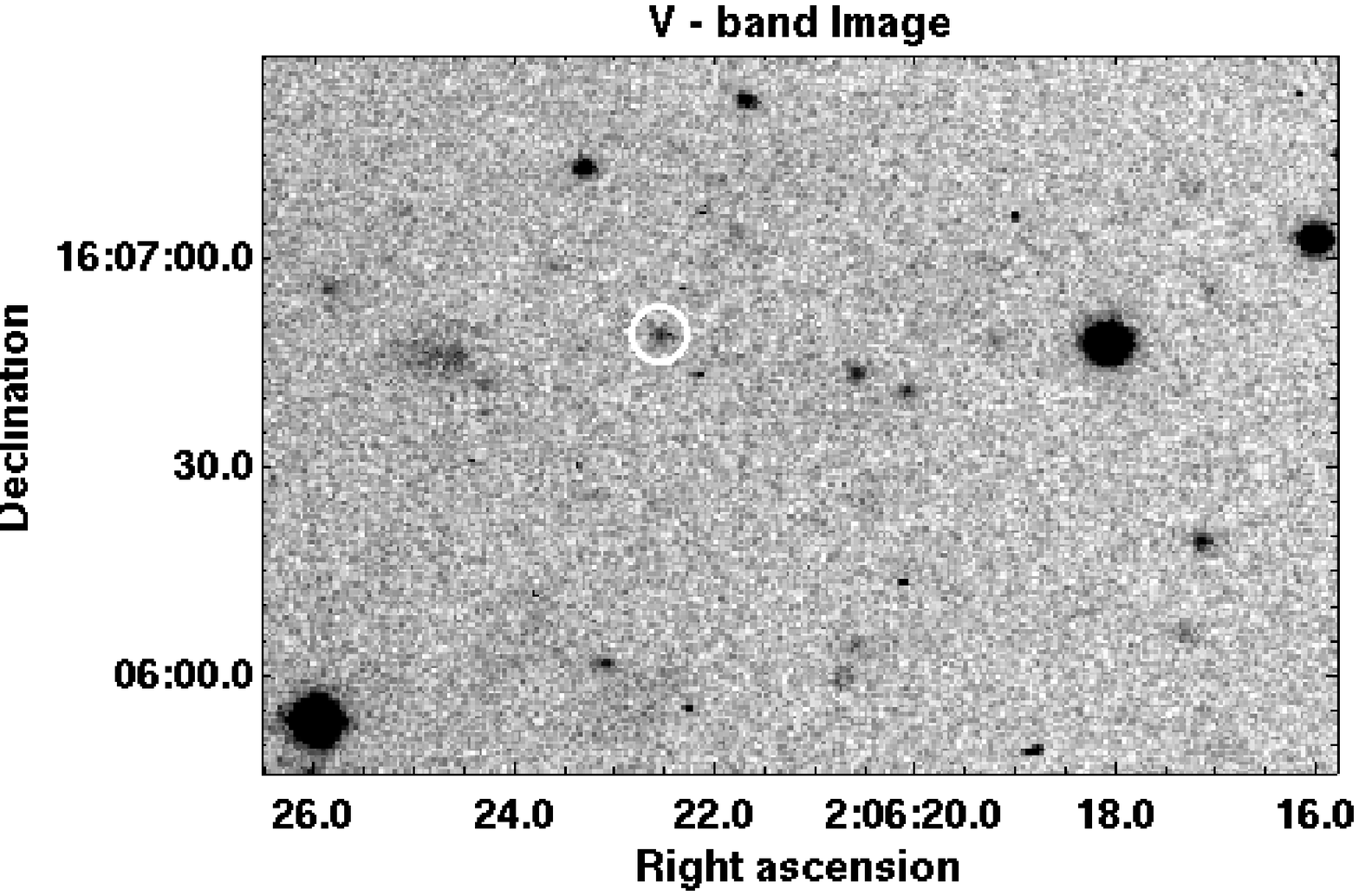}
\caption{The $R$-band and $V$-band images of GRB 040924 taken with LOT.
The location of the afterglow is indicated by a circle in each image.}
\label{grb040924image}
\end{figure}

\clearpage

\begin{figure}
\plotone{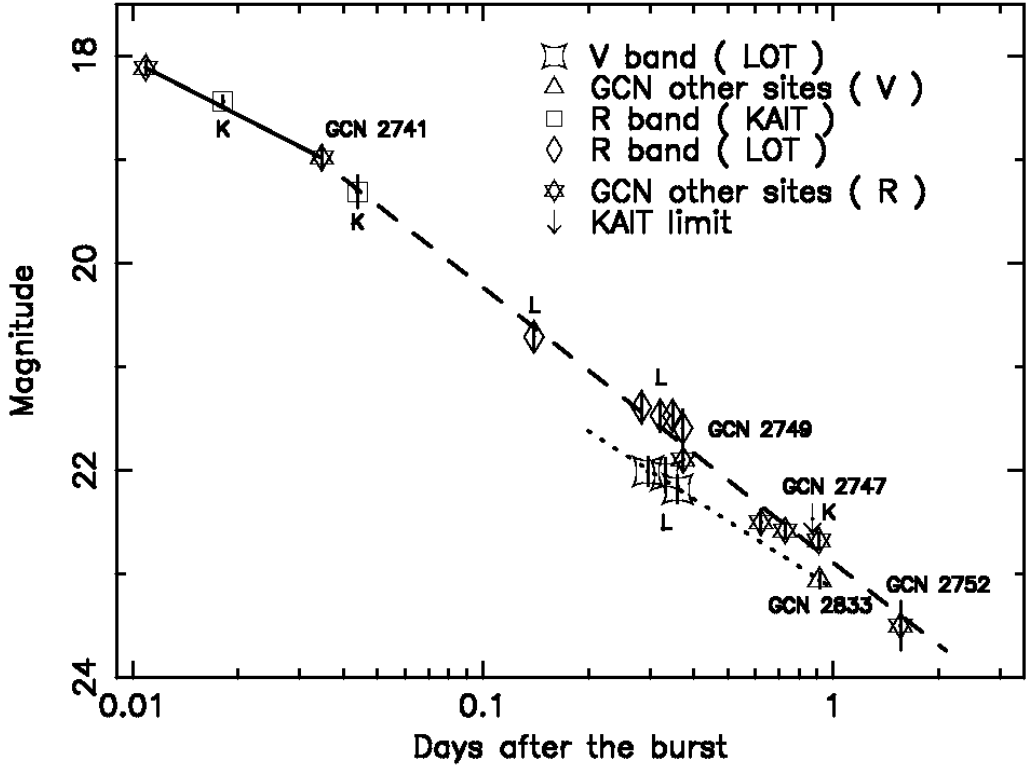}
\caption{The $V$-band and $R$-band light curves based on our (LOT
and KAIT) observations and the recalibrated data points of Fox (2004),
Fynbo et al. (2004), Khamitov et al. (2004b,c), and Silvey et
al. (2004). The straight lines represent the power-law models [$F(t)
\propto t^{\alpha}$] fitted to the data points: {\it solid} is for the
$R$-band $\alpha = -0.7$ at early times (based on the first three
observations), {\it dashed} is the
late-time $R$-band best fit of $\alpha = -1.06 \pm 0.03$ starting from
the third observation, and {\it dotted} is the $V$-band best fit of
$\alpha = -0.87 \pm 0.02$ from LOT and Silvey et al. (2004).}
\label{grb040924lc}
\end{figure}

\clearpage

\begin{table}[htb]
\begin{center}
\caption{Log of GRB 040924 Optical Afterglow Observations}
\begin{tabular}{ccccccc}
\tableline\tableline
UT Date & Start Time & Mean Delay (days) & Filter  & Exposure (s) & mag & Site\\
\tableline
2004-09-24 & 18:52:37 &  0.296 & $V$  & 300 s $\times$ 3 & 22.01$\pm$0.13 & LOT \\  
2004-09-24 & 19:46:02 &  0.333 & $V$  & 300 s $\times$ 3 & 22.05$\pm$0.16 & LOT \\  
2004-09-24 & 20:24:54 &  0.360 & $V$  & 300 s $\times$ 3 & 22.18$\pm$0.13 & LOT \\ \hline
2004-09-24 & 12:18:21 &  0.018 & $R$ & 120 s $\times$ 1 & 18.44$\pm$0.05 & KAIT\tablenotemark{a} \\
2004-09-24 & 12:55:21 &  0.044 & $R$ & 120 s $\times$ 1 & 19.31$\pm$0.15 & KAIT\tablenotemark{a} \\
2004-09-24 & 15:00:55 &  0.140 & $R$ & 600 s $\times$ 2 & 20.71$\pm$0.13 & LOT \\
2004-09-24 & 18:34:37 &  0.284 & $R$ & 300 s $\times$ 3 & 21.39$\pm$0.15 & LOT \\
2004-09-24 & 19:28:57 &  0.321 & $R$ & 300 s $\times$ 3 & 21.47$\pm$0.15 & LOT \\  
2004-09-24 & 20:07:48 &  0.348 & $R$ & 300 s $\times$ 3 & 21.47$\pm$0.14 & LOT \\  
2004-09-24 & 20:42:17 &  0.372 & $R$ & 300 s $\times$ 3 & 21.59$\pm$0.17 & LOT \\
2004-09-25 & 08:35:00 &  0.873 & $R$  & 300 s $\times$ 3 & $>$22.47    & KAIT\tablenotemark{a} \\  
\tablenotetext{a}{KAIT measurements were unfiltered, but transformed to $R$ (Li et al. 2003a,b).}
\end{tabular}
\end{center}
\end{table}

\clearpage

\begin{table}[htb]
\begin{center}
\caption{Electron Spectral Index ($p$) Calculated from the
Measured Spectral Index ($\alpha$)}
\vspace{0.3cm}
\begin{tabular}{ccccccccc}
\tableline\tableline & & \multicolumn{1}{r}{$p > 2$}& & &
&&\multicolumn{1}{r}{$1 < p < 2$} &\\ \cline{3-5} \cline{7-9}
\multicolumn{1}{c}{\raisebox{1.5ex}{Frequency\tablenotemark{a}}}&
\multicolumn{1}{c}{\raisebox{1.5ex}{model}}&
\multicolumn{1}{c}{relation\tablenotemark{b}}&
\multicolumn{1}{c}{$p_{1}$\tablenotemark{c}}&
\multicolumn{1}{c}{$p_{2}$\tablenotemark{c}}& \multicolumn{1}{c}{}&
\multicolumn{1}{c}{relation\tablenotemark{b}}&
\multicolumn{1}{c}{$p_{1}$\tablenotemark{c}}&
\multicolumn{1}{c}{$p_{2}$\tablenotemark{c}}\\ \tableline $\nu_{\rm opt}<
\nu_{\rm c}$ & ISM &$p=1-4\alpha/3$
& 1.93 & 2.41 & & $p=-2\alpha-10/3$ & $-1.93$ & $-1.21$ \\ $\nu_{\rm opt}> \nu_{\rm c}$ &
ISM & $p=2/3-4\alpha/3$ & 1.60 & 2.08 & & $p=-2-16\alpha/3$ & ~1.73 & ~3.65\\
\tableline $\nu_{\rm opt}< \nu_{\rm c}$ & Wind & $p=1/3-4\alpha/3$
& 1.26 & 1.74 & & $p=-6-8\alpha$ & $-0.4$ & ~2.48\\ $\nu_{\rm opt}> \nu_{\rm c}$ &
Wind & $p=2/3-4\alpha/3$ & 1.6 & 2.08& &$p=-8-8\alpha$ & $-2.4$ & ~0.48 \\
\tableline $\nu_{\rm opt}< \nu_{\rm c}$ & Jet & $p=-\alpha$
& 0.7 & 1.06& & $p=-6-4\alpha$ & $-3.2$ & $-1.76$ \\ $\nu_{\rm opt}> \nu_{\rm c}$ &
Jet & $p=-\alpha$ & 0.7 & 1.06 & & $p=-6-4\alpha$ & $-3.2$ & $-1.76$ \\
\tablenotetext{a}{The frequency at which the spectrum breaks due to
synchrotron cooling is $\nu_{\rm c}$, whereas the typical visible-light
frequency is $\nu_{\rm opt}$.}  \tablenotetext{b}{The GRB afterglow model
relation of Zhang \& M\'esz\'aros (2004).}  \tablenotetext{c}{The
electron spectral index calculated from $\alpha_1 = -0.70$ and
$\alpha_2 = -1.06$.}
\end{tabular}
\end{center}
\end{table}

\end{document}